# Solution to the hole-doping problem and tunable quantum Hall effect in Bi$_2$Se$_3$ thin films

*Jisoo Moon[1], Nikesh Koirala[1,†], Maryam Salehi[2], Wenhan Zhang[1], Weida Wu[1], Seongshik Oh[1*]*

[1]Department of Physics & Astronomy, Rutgers, The State University of New Jersey, Piscataway, New Jersey 08854, U.S.A.

[2]Department of Materials Science & Engineering, Rutgers, The State University of New Jersey, Piscataway, New Jersey 08854, U.S.A.

*Correspondence should be addressed to ohsean@physics.rutgers.edu and +1 (848) 445-8754 (S.O.)

ABSTRACT. $Bi_2Se_3$, one of the most widely studied topological insulators (TIs), is naturally electron-doped due to n-type native defects. However, many years of efforts to achieve p-type $Bi_2Se_3$ thin films have failed so far. Here, we provide a solution to this long-standing problem, showing that the main culprit has been the high density of interfacial defects. By suppressing these defects through an interfacial engineering scheme, we have successfully implemented p-type $Bi_2Se_3$ thin films down to the thinnest topological regime. On this platform, we present the first tunable quantum Hall effect (QHE) study in $Bi_2Se_3$ thin films, and reveal not only significantly asymmetric QHE signatures across the Dirac point but also the presence of competing anomalous states near the zeroth Landau level. The availability of doping tunable $Bi_2Se_3$ thin films will now make it possible to implement various topological quantum devices, previously inaccessible.

Hole (p) doping has been challenging in $Bi_2Se_3$[1–3], one of the most widely studied topological insulators (TIs)[4–10]. Unlike conventional semiconductor materials, the problem is complicated due to the presence of both surface and bulk states in topological insulators: we have to consider the doping problem of the surface and the bulk states separately. Both the surface and bulk states of $Bi_2Se_3$ have a strong tendency toward n-type due to its native n-type defects such as selenium vacancies[1–3,11–17]. In bulk crystals compensation dopants such as Ca and Mn can be used to convert the dominant carrier type from n- to p-type[18–25]. However, such a compensation doping scheme has not been successful in thin films of $Bi_2Se_3$. More specifically, we have tried various potential p-type dopants such as Zn, Mg, Ca, Sr, and Ba as compensation dopants, but none of them have so far led to p-type $Bi_2Se_3$ thin films; no p-type $Bi_2Se_3$ thin films have been demonstrated in the literature either. Only when they were quite thick (~200 nm), we were able to achieve p-type $Bi_2Se_3$ films through a complex process[8]. This difficulty in achieving p-type $Bi_2Se_3$ thin films has been puzzling, considering the very existence of p-type $Bi_2Se_3$ bulk crystals[19,20]. It may be suspected that this discrepancy in doping efficiency between thin films and bulk crystals could be due to the different growth conditions of the two systems, such as growth temperatures, considering that films are, in general, grown at much lower temperatures than bulk crystals. However, it should be noted that even p-type $Bi_2Se_3$ bulk crystals tend to become n-type when the crystals are made into thin flakes through, say, the Scotch tape method[26]. $Bi_2Se_3$ flakes can remain p-type only if they are relatively thick (> ~150 nm)[24]. All these observations provide evidence that whether they are thin films or bulk crystals, thickness critically affects the doping efficiency of the $Bi_2Se_3$ system, but the origin behind this problem has been a mystery, not to mention that a solution does not exist. Finding a solution to this p-doping mystery in $Bi_2Se_3$ thin films would be a major breakthrough toward the age of topological electronics, just as the solution to the long-standing p-doping problem in GaN has led to the solid-state-lighting revolution[27].

In order to better understand the hole-doping problem, we grew a series of $Bi_2Se_3$ thin films with different compensation doping schemes and different thicknesses and studied their carrier types and densities as shown in Figure 1. As for the compensation dopant, we have chosen Ca, but according to our comparison studies (Supporting Information Figure S3 and Section IV), other alkaline earth elements such as Mg and Sr might work as well. Figure 1a shows that when $Bi_2Se_3$ thin films are directly grown on $Al_2O_3$ (0001) substrates without any capping layer or a buffer layer, the films remain n-type regardless of the compensation-doping-level for film thickness up to 50 QL (quintuple layer; 1 QL is approximately 1 nm thick): as for how to control and estimate the Ca-doping level, see Supporting Information Section II.

In order to check the effect of the air exposure, we subsequently grew another set of samples with Se capping layer, which was previously shown to be effective in protecting the surface of $Bi_2Se_3$ films from air exposure[28]. As depicted in Figure 1b, the Se capping does make a difference and 50 QL $Bi_2Se_3$ film on $Al_2O_3$ is now converted to p-type. This suggests that part of the reason why counter-doping for $Bi_2Se_3$ thin films becomes ineffective is due to air exposure, which could neutralize the easily-oxidizing Ca dopant as well as add extra n-type dopant such as water vapor[10]. However, even with the Se capping, the counter-doping still fails for much thinner (6 QL) films. This shows that air exposure is not the only reason for the failure of charge compensation doping with Ca, suggesting that the interface with the substrate is another potential source of the problem.

As reported recently[6], most of the native defects responsible for the high level of n-type carrier densities in $Bi_2Se_3$ thin films originate from the interface with commercially available substrates due to chemical and structural mismatch. This leads to high density of n-type carriers and prevents $Bi_2Se_3$ thin films from being hole-doped[29]. Therefore, we employ $In_2Se_3$ as a buffer layer that works as an excellent substrate for $Bi_2Se_3$ thin films, sharing chemical similarity and the same crystal structure. However, $In_2Se_3$ has multiple phases unlike $Bi_2Se_3$, and grows on $Al_2O_3$ in a disordered form. In order to overcome this problem, we grow a seed layer of 3 QL $Bi_2Se_3$ on $Al_2O_3$ substrate at 135 °C before growing $In_2Se_3$: for all subsequent layers, our typical growth rate is ~1 QL/min. The seed layer serves as a substrate for $In_2Se_3$, and it is followed by 20 QL of $In_2Se_3$ deposition at 300 °C. Then, we anneal it to 600 °C, which forces the $Bi_2Se_3$ layer to be diffused into $In_2Se_3$ layer and eventually vaporized out. As the last step in the buffer layer growth, 20 QL $(Bi_{0.5}In_{0.5})_2Se_3$ (in short BIS) is deposited on the highly crystalline $In_2Se_3$ layer at 275 °C. This 50% BIS layer provides $Bi_2Se_3$ not only with less lattice mismatch, ~1.6 %, than $In_2Se_3$ layer (~3.3%) but also with suppression of In diffusion into the (Ca-doped) $Bi_2Se_3$ film on top. Now, we grow Ca-doped $Bi_2Se_3$ film on the BIS at 275 °C, then an atomically sharp and defect-free interface between $Bi_2Se_3$ and the BIS can be obtained. Utilizing the properly-designed BIS buffer layer, we were able to eliminate most of these interfacial defects. The surface Fermi levels of these films were so low that they revealed a series of - previously inaccessible - topological quantum effects such as quantum Hall effect (QHE)[6], quantized Faraday/Kerr rotation[30], and quantum size effect during topological phase transitions[31]. The significantly reduced defect density in these films can be seen in the Hall resistance data of Figure 1c: an 8 QL $Bi_2Se_3$ thin film directly grown on $Al_2O_3$ (0001) shows an n-type sheet carrier density of $2.5 \times 10^{13}$ /cm$^2$, whereas a similar film on the BIS buffer layer exhibits only $1.0 \times 10^{12}$ /cm$^2$, 1/25 of the former value.

Now, by employing both the BIS buffer layer and the Se capping layer, we were able to implement p-type $Bi_2Se_3$ thin films down to the thinnest topological regime (6 QL) as shown in Figure 1d. As summarized in Figure 1e, previous failures of counter-doping for thin $Bi_2Se_3$ films originate from two factors: first, neutralization of the counter-dopants due to air exposure on the top surface and second, high density of defects at the interface with the substrate. When $Bi_2Se_3$ films are grown directly on commercial substrates, there exists a high density of defects confined to the interface between the film and the substrate. These interfacial defects lead to high level of n-type carriers with the surface Fermi level located far above the bottom of the conduction band minimum[2,8,32]. Our current study suggests that this interfacial defect density on the commercial substrate is higher than the solubility limit of the compensation dopant so that the film remains n-type up to the maximum counter-doping (Figure 1f). On the other hand, with the BIS buffer layer, the interfacial defect density drops below the solubility limit of the compensation dopant and the Fermi level can be tuned below the Dirac point, leading to p-type $Bi_2Se_3$ films.

We now provide more detailed Hall effect data for a range of Ca-doped $Bi_2Se_3$ films grown on the BIS buffer and the Se capping layers in Figure 2a. As the Ca doping level increases and compensates for the intrinsic n-type carriers, the (n-type) sheet carrier density gradually decreases, which appears as increasing negative slope of the $R_{xy}(B)$ curve as illustrated in Figure 2c. i-ii. As we add more Ca to the films, the slope changes from negative (n-type) to positive (p-type), passing through a nonlinear n-p mixed regime: Figure 2c. iii-iv. 50 QL films enter the n-p mixed regime at 0.09% of Ca and become fully p-type at 0.13%. However, as the films get thinner, gradually higher level of Ca doping is required to reach the n-p mixed and the p regimes. The n-p mixed regime started at 0.13% for 20 QL, 0.3% for 10 QL, and 0.9% for 6 QL, and the p regime started at 0.3% for 20 QL, 2.0% for 10 QL and 6 QL. As summarized in the phase diagram in Figure 2b, these observations show that counter-doping becomes gradually more difficult, requiring a higher level of Ca doping, as Bi$_2$Se$_3$ films become thinner. This suggests that even in these interface-engineered Bi$_2$Se$_3$ films, the majority of defects still originate from the interfaces rather than from the bulk of the films.

Even though p-type Bi$_2$Se$_3$ films are achieved for all thicknesses from 50 QL down to 6 QL at 2% of Ca, the films again become n-type with further increase (above 4.4%) in Ca doping except for 6 QL, which becomes insulating probably due to a disorder-driven topological phase transition[31,33]. This is because Ca doping inevitably introduces disorder and thus degrades both electronic and structural properties of the Bi$_2$Se$_3$ films. This can be seen in the carrier density and mobility vs Ca doping plots of Figure 3a and b. In Figure 3a, we plot the sheet carrier density ($n_{2D}$) vs Ca doping for the 20 QL Bi$_2$Se$_3$ films grown on the BIS buffer with Se capping layers. For simplicity, we used the Hall resistance data at $B = 0$ T for the carrier density, and excluded the nonlinear n-p mixed regime because of ill-defined $n_{2D}$ in nonlinear $R_{xy}(B)$ curves: more elaborate discussion of the data including the n-p mixed regime is given in Supporting Information Section III. It shows that the carrier type changes from n to p, and finally to n again at very high Ca concentration. In Figure 3b, we plot the corresponding mobility calculated from $\mu = 1 / (n_{2D} \cdot e \cdot R_{xx})$, where $R_{xx}$ is the sheet resistance. The mobility shows a sharp drop as soon as the film becomes p-type, and gradually degrades with increasing Ca concentration in the p-regime: as we further discuss with Figure 4d below, the sharp drop of the mobility through the charge neutral point seems to be an intrinsic property of the Bi$_2$Se$_3$ band structure, whereas its gradual decrease at higher Ca concentration is more due to disorder.

This observation suggests that once Ca concentration goes beyond a certain solubility limit, Ca in Bi$_2$Se$_3$ stops acting as a p-type dopant and instead starts to act as n-type. This is well understood from the very fact that most defects in $Bi_2Se_3$ behave as n-type[2]. High level of Ca doping degrades not only the mobility but also the morphology, as can be seen in the Scanning Tunneling Microscope (STM) images of Figure 3c and d. The STM images show that the morphology of $Bi_2Se_3$ films degrades noticeably at high Ca concentration. The re-entrant n-type behavior beyond a solubility limit of Ca doping explains why $Bi_2Se_3$ thin films without the buffer layer cannot be turned into p-type via compensation doping: high density of (n-type) interfacial defects requires proportionately high density of compensation dopant to reach p-type, but beyond a certain doping density, the compensation dopant starts to act as n- instead of p-type, making it impossible to convert the film into p-type.

Finally, these p-type $Bi_2Se_3$ thin films have allowed us to take the first doping-dependent QHE study in TI materials. Although gate-dependent QHE have been observed in $(Bi_{1-x}Sb_x)_2Te_3$ thin films[34] and in $BiSbTeSe_2$ single crystal flakes[35], and non-gated QHE in $Bi_2Se_3$ thin films[6], doping-dependent QHE has not been available in any TI materials. Getting access to quantum regime in both n and p-doped TI films is critical for many topological quantum devices. For this study, we grew a series of 8 QL $Bi_2Se_3$ thin films with different Ca doping levels and measured their Hall effect up to a 34.5 T magnetic field with hand-patterned Hall bar geometry. In addition, we added $MoO_3$ as the main capping layer under the Se capping as shown in Figure 4d, because $MoO_3$ helps reduce the Ca doping level to reach the p-type[6]. Figure 4a shows Ca-doping-dependent QHE in the $Bi_2Se_3$ films. The Hall resistance at high magnetic field is perfectly quantized at $R_{xy} = h/e^2 = 25.8$ kΩ with vanishing sheet resistance on the n-side up to 0.08% of Ca, indicating the emergence of a well-defined chiral edge channel at high magnetic fields. Here, it is notable that only the 1st Landau level (LL) plateau ($R_{xy} = h/(ve^2)$ with $v = 1$) is observable. The absence of higher level plateaus could be due to multiple factors such as the formation of non-chiral edge channels on the side surfaces[36], the presence of electronic puddles[11,37] and non-zero coupling between the transport channels of the top and bottom surface states[4].

As Ca doping is increased toward the Dirac point (or, equivalently, zeroth LL), while the Hall resistance gradually deviates from the quantized value, the sheet magnetoresistance changes drastically even with minute change in doping. At 0.09%, the Hall resistance at high magnetic field deviates just slightly from the quantized value, but the sheet magnetoresistance soars from close-to-zero to a large insulating value (~40 kΩ) at high magnetic fields as the doping changes from 0.08 to 0.09% even if the resistance remains nearly the same at zero magnetic field. This nominally implies that the dissipationless edge state is breaking down while the rest of the sample is insulating under high magnetic field. However, as we get even closer to the zeroth LL at 0.11%, the Hall resistance starts to decrease over the entire magnetic field but the sheet magnetoresistance plummets: this nominally implies that the - otherwise insulating - sample is now becoming conducting. The overall decrease in Hall resistance at 0.11% is likely due to the n-p mixing as we discussed above in Figure 2. On the other hand, the presence of both insulating and conducting - in terms of sheet magnetoresistance - states right next to each other near the zeroth LL is puzzling, and it suggests that the zeroth LL harbors multiple unidentified states, which would require in-depth future studies, both experimentally and theoretically.

With further increase of Ca doping to 0.12 and 0.14%, the majority carrier type now converts to p-type. There are a couple of features noticeable in the p-type films. First, the p-type carrier density grows quickly to ~$10^{12}$ /cm$^2$ (Figure 4b) with a noticeable signature of n-p mixing: pure p-regime can be achieved with much higher (~1%) level of Ca doping, but only with about an order of magnitude higher carrier density than the pure n-regime (see Supporting Information Section V). Second, the signature of QHE is significantly degraded in the p-regime with the 1st

LL barely observable only for 0.12%. Third, as can be seen in Figure 4c, the mobility sharply decreases as soon as the majority carrier type changes from n to p-type: this is consistent with what we observed in Figure 3. Although Ca-doping can degrade the mobility to some extent, it cannot be the main factor at these low doping levels: note that the mobility in Figure 4c remains almost constant or even grows a little with Ca-doping in the n-regime until it reaches the charge neutral point, after which the mobility sharply decreases. This suggests that the mobility drop right on the p-side must be of intrinsic origin. Both the sharp drop in mobility and the sharp increase in carrier density on the p-side are very likely due to the band structure effect. According to the band diagram of $Bi_2Se_3$[38], the Dirac point is very close to the bulk valence band and the surface band broadens noticeably on the p-side: both of these effects can account for the observed degradation of transport properties on the p-side of $Bi_2Se_3$ thin films. The near-absence of QHE and low mobility on the p-side of $Bi_2Se_3$ system is also consistent with the absence of Landau levels on the p-side as measured by a previous Scanning Tunneling Spectroscopy (STS) study[39].

In conclusion, we have provided a solution to the long-standing hole doping problem and presented the first doping-dependent QHE study in $Bi_2Se_3$ thin films. We have shown that the previous failures of hole doping efforts in $Bi_2Se_3$ films are due to: first, large density of interfacial defects from the substrates, and second, neutralization of the dopant from the air. By suppressing the interfacial defects via the $(Bi_{0.5}In_{0.5})_2Se_3$ buffer layer and protecting the top surface with a capping layer, we were able to reach the hole regime of $Bi_2Se_3$ films down to the thinnest topological regime (6 QL) through compensation doping. In addition, in the doping dependent QHE study of 8 QL $Bi_2Se_3$ thin films we showed that the first (and only) Landau level is clearly observed on the electron side but quickly dissolves away on the hole side. Interestingly, at close proximity to the zeroth Landau level, we observed both the disintegration of the edge channel and the emergence of a conducting state, whose origins are unknown at the moment. Altogether, just as the solution to the p-doping problem of GaN has led to the LED lighting revolution, we believe that this solution to the p-doping problem of $Bi_2Se_3$ thin films will help usher in the age of topological quantum electronics.

**MBE growth**

Films were grown on 10 mm × 10 mm $Al_2O_3$ (0001) substrates (MTI Corporation) using a custom-built MBE system (SVTA) with base pressure less than $5 \times 10^{-10}$ Torr. Substrates were cleaned ex situ by UV-generated ozone and in situ by heating to 750 ºC under oxygen pressure of $1 \times 10^{-6}$ Torr. High purity (99.999%) Bi, In, Se, and Ca were thermally evaporated using Knudsen effusion cells for the film growth. Flux ratio of Se to Bi and In was maintained at above 10:1 to prevent the films from defects arising from Se deficiency. Source fluxes were calibrated in situ by quartz crystal micro-balance (QCM). For growth of $In_2Se_3$-$(Bi_{0.5}In_{0.5})_2Se_3$ buffer layer and control of Ca flux relative to Bi (doping level), see Supporting Information Section I and II. $MoO_3$ (for 8QL films) and Se were deposited at room temperature for capping on top of $(Ca_xBi_{1-x})_2Se_3$ layers.

**Transport measurements**

DC magneto-resistance and the Hall resistance measurements were carried out in a liquid He cryogenic system (AMI) with base temperature of 1.6 K. Perpendicular magnetic field (B) was applied to the films up to ± 9 Tesla using a solenoid superconductor coil. Electrical contacts were made by pressing thin In leads in van der Pauw geometry. Electrical current was generated and longitudinal, transverse voltages were measured for $R_{xx}$ and $R_{xy}$ using KE2400 sourcemeter and KE7001 switch matrix system. Raw data of $R_{xx}(B)$ and $R_{xy}(B)$ were symmetrized and antisymmetrized, respectively.

**STM measurements**

STM measurements were carried out at 77 K in an Omicron LT-STM with a base pressure of $8 \times 10^{-12}$ Torr. Electrochemically etched tungsten tip was characterized on single crystal Au (111) surface. The samples, which were capped by Se in the MBE chamber, were exposed to air for several minutes during transfer to the STM chamber and the Se capping was decapped in a prep chamber, attached to the STM main chamber, with a base pressure of $2 \times 10^{-11}$ Torr. The samples were first sputtered with 500 eV argon ions for 15 minutes. Subsequently, they were annealed up to ~200 °C to desorb the amorphous Se, after which they were immediately transferred into the STM head for measurements[40].

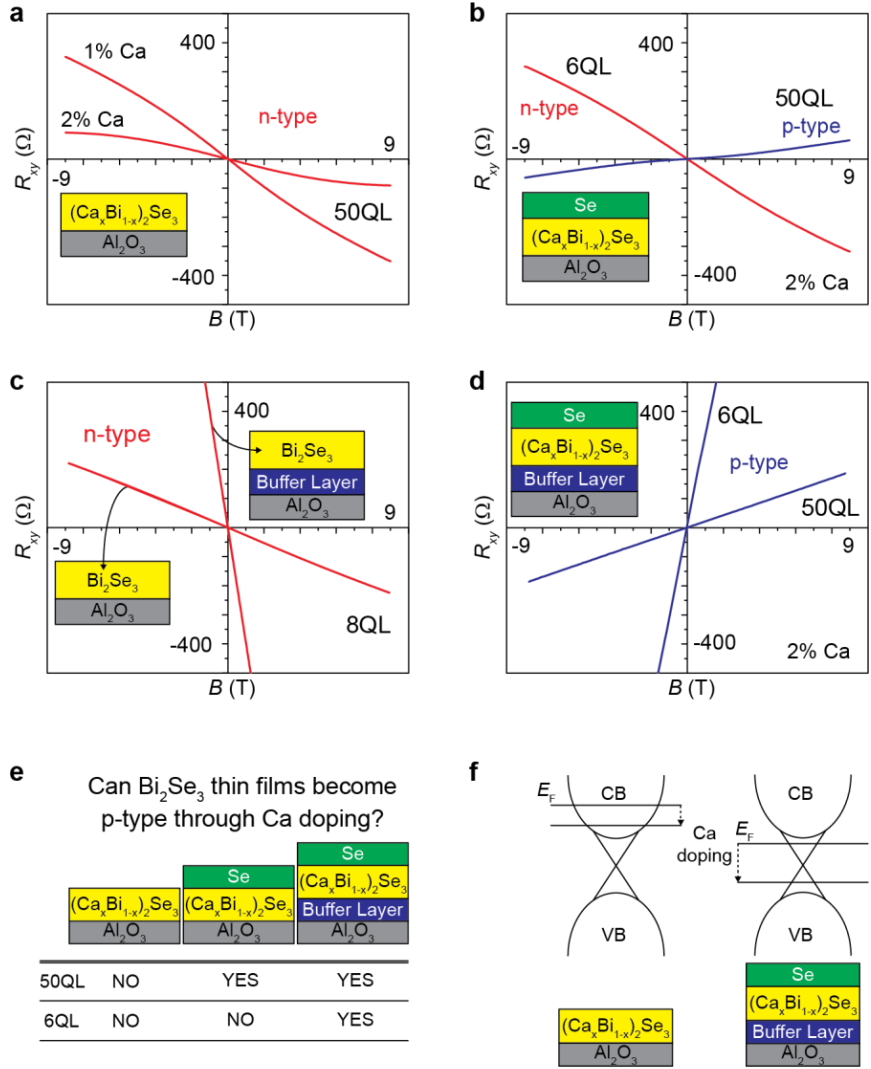

**Figure 1.** Ca doping study in $Bi_2Se_3$ thin films with different interfacial layers (at 1.5 K). (a-d) Hall resistance data, (e) summary table, and (f) schematic of energy band diagram with and without the interface layers. (a) Grown directly on $Al_2O_3$ (0001) substrates: the films remain n-type regardless of Ca-doping levels. (b) Grown on $Al_2O_3$ substrates with a Se capping layer: thick (50 QL) film can become p-type whereas thin (6 QL) film still remains n-type. (c) Effect of the BIS buffer layer on the interfacial defect density: the $Bi_2Se_3$ film grown on the buffer layer shows much lower carrier density than the one directly grown on $Al_2O_3$. (d) Achievement of p-type $Bi_2Se_3$ films down to 6 QL with the BIS buffer and the Se capping layers. (e) Summary table for the counter-

doping effectiveness for different thicknesses and interfacial layers. (f) Schematic drawings showing the Fermi level tunability with Ca doping in Bi$_2$Se$_3$ films grown with and without the interfacial layers.

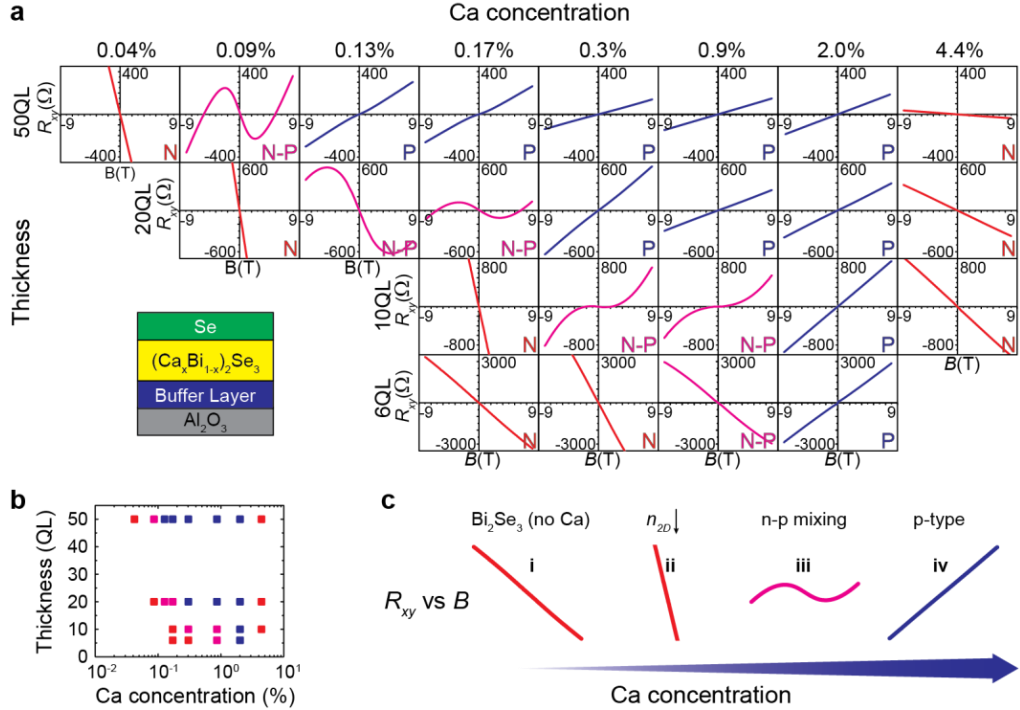

**Figure 2.** Hall resistance study for n- and p-type Bi$_2$Se$_3$ thin films (at 1.5 K). (a) $R_{xy}(B)$ data vs Ca concentration and thickness for Ca-doped Bi$_2$Se$_3$ films with the BIS buffer and the Se capping layers. n-type (negative slope) and p-type (positive slope) curves are colored as red and blue, respectively, and the nonlinear, n-p mixed curves are colored as pink. As Ca concentration increases, all the films transition from n- to p-type through n-p mixed regime, and then eventually become n-type again except for the 6 QL film, which becomes insulating, instead. (b) A phase diagram for the carrier type in the thickness and Ca concentration space: n, n-p mixed, and p regime are depicted as red, pink and blue squares, respectively. Note that the phase space for the p regime

shrinks as the film gets thin. (c) A cartoon showing the common trend of $R_{xy}(B)$ depending on Ca concentration for all thicknesses. (i) $Bi_2Se_3$ is naturally n-type without Ca dopants. (ii) As Ca doping level increases, the (n-type) sheet carrier density gradually decreases, resulting in increased negative slope of the $R_{xy}(B)$ curve. (iii-iv) With more Ca doping, the slope changes from negative (n-type) to positive (p-type) passing through a nonlinear n-p mixed regime.

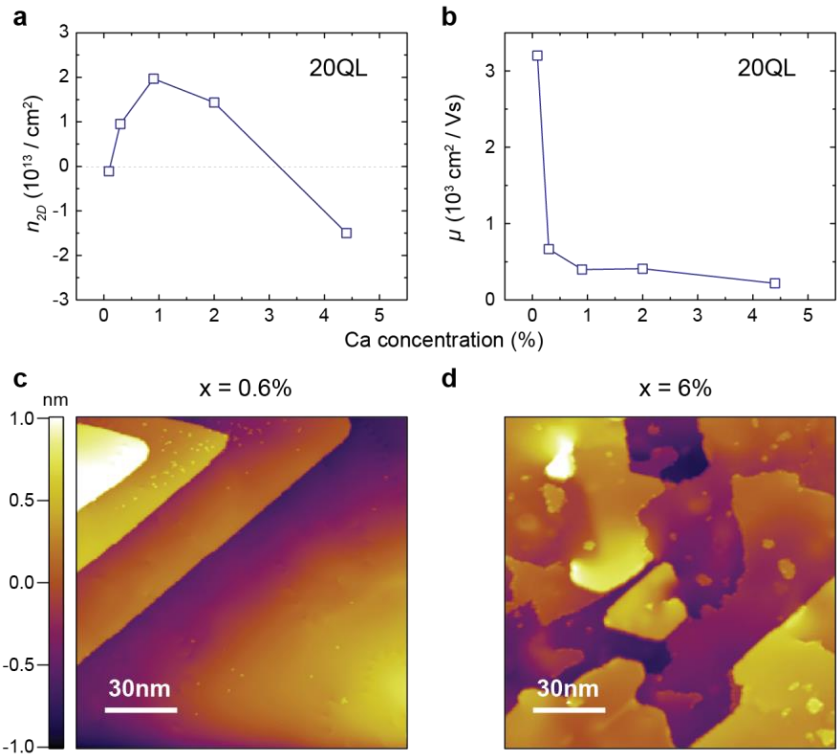

**Figure 3.** Transport properties and morphology comparison in Ca-doped $Bi_2Se_3$ films. (a) 2D sheet carrier density ($n_{2D}$) and (b) mobility ($\mu$) vs Ca concentration of the 20 QL, $Bi_2Se_3$ films on BIS buffer layer with Se capping. The carrier densities are taken from the 20 QL Hall effect data of Figure 2 at $B = 0$ T. Data from severely nonlinear $R_{xy}(B)$ curves (0.13% and 0.17% of Ca) are excluded due to their uncertain carrier densities. The carrier type changes from n to p, and finally

to n, while the mobility decreases sharply right around the n-p crossing point and gradually as the Ca concentration increases further. (c, d) STM topography images of Ca-doped $Bi_2Se_3$ films at (c) x = 0.6% and (d) x = 6%. The morphology degrades significantly at the higher Ca concentration, where the carrier type of the film reenters into the n regime from the p.

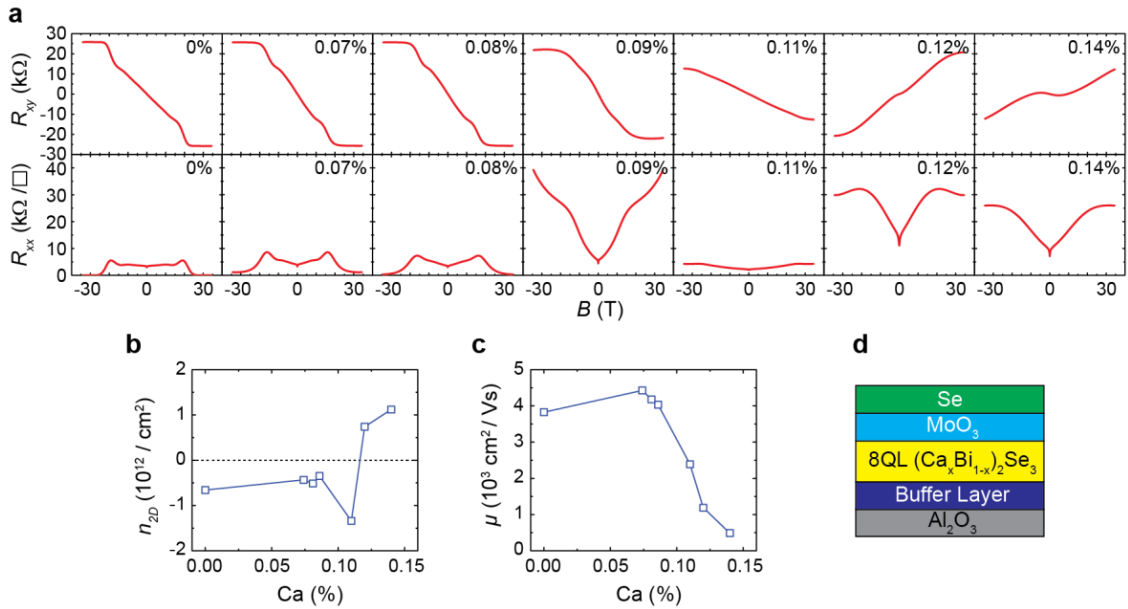

**Figure 4.** Ca doping-dependent transport properties, around the charge neutral point, of 8 QL $Bi_2Se_3$ films capped by $MoO_3$ and Se (at 300 mK). See Supporting Information Figure S4 and Section V for a broader doping dependence of a similar set of films. (a) Magnetic field dependence of Hall ($R_{xy}$) and sheet ($R_{xx}$) resistance with different Ca doping levels. (b) 2D sheet carrier density ($n_{2D}$) and (c) mobility ($\mu$) with different Ca doping levels. We used $(dR_{xy}(B)/dB)_{B=0T}$ for 0% ~ 0.11%, and $(dR_{xy}(B)/dB)_{B=15, 20T}$ for 0.12% and 0.14% respectively, to extract net carrier density and mobility from the nonlinear Hall curves while avoiding the n-p mixing and the quantum Hall effects. Note that the mobility sharply decreases when the carrier type changes from n to p. (d)

Schematic layer structure of the 8 QL Ca-doped $Bi_2Se_3$ films grown on the BIS buffer layer and capped by $MoO_3$ and Se. The pure $Bi_2Se_3$ data (0% of Ca, the leftmost data in a-c) are taken from Ref.[6].

ACKNOWLEDGMENT

We thank N. Peter Armitage for comments and Pavel P. Shibayev for proofreading. This work is supported by Gordon and Betty Moore Foundation's EPiQS Initiative (GBMF4418) and National Science Foundation (NSF) (EFMA-1542798). Transport measurements in National High Magnetic Field Laboratory were further supported by NSF (DMR-1157490) and the State University of Florida.

# Supporting Information

# Solution to the hole-doping problem and tunable quantum Hall effect in $Bi_2Se_3$ thin films

Jisoo Moon[1], Nikesh Koirala[1,†], Maryam Salehi[2], Wenhan Zhang[1], Weida Wu[1], Seongshik Oh[1*]

[1]Department of Physics & Astronomy, Rutgers, The State University of New Jersey, Piscataway, New Jersey 08854, U.S.A.

[2]Department of Materials Science & Engineering, Rutgers, The State University of New Jersey, Piscataway, New Jersey 08854, U.S.A.

*Correspondence should be addressed to ohsean@physics.rutgers.edu and +1 (848) 445-8754 (S.O.)

†Current address: Department of Physics, Massachusetts Institute of Technology, Cambridge, MA 02139, U.S.A.

**Content**

## I. Characterization of In$_2$Se$_3$-(Bi$_{0.5}$In$_{0.5}$)$_2$Se$_3$ Buffer Layer

As mentioned in the main text, growth of Bi$_2$Se$_3$ thin films directly on commercial substrates leaves high density of interfacial defects due to chemical and structural mismatch. This leads to high density of n-type carriers and prevents Bi$_2$Se$_3$ thin films from being hole-doped[1]. Therefore, the In$_2$Se$_3$-(Bi$_{0.5}$In$_{0.5}$)$_2$Se$_3$ buffer layer (BIS-BL) is employed to solve this problem and make the hole doping of Bi$_2$Se$_3$ thin films possible. Below is detailed information of the BIS-BL and the Bi$_2$Se$_3$ thin films grown on the BIS-BL. Full details can be found in Ref. 2 and its Supporting Information.

Bi$_2$Se$_3$ films grown on BIS-BL exhibit much higher crystalline quality than those on commercial substrates. This is shown by reflection high-energy electron diffraction (RHEED) patterns and (scanning) transmission electron microscopy ((S)TEM) images in Fig. S1. The RHEED streaks of Bi$_2$Se$_3$ grown on BIS-BL are much brighter than Bi$_2$Se$_3$ directly grown on Al$_2$O$_3$ (0001) substrate as shown in the 2$^{nd}$ and 6$^{th}$ columns in Fig. S1a. The brighter RHEED pattern in reciprocal space indicates lower disordered atomic surface structure in real space. In addition, STEM images of Bi$_2$Se$_3$ grown on BIS-BL in Fig. S1b and c show atomically sharp interface between Bi$_2$Se$_3$ and BIS layers. Compared to the images of BIS-BL, Bi$_2$Se$_3$ films grown on Al$_2$O$_3$ (0001) and Si (111) in Fig. S1d and e show disordered interface between Bi$_2$Se$_3$ and the commercial substrates. Altogether, the Bi$_2$Se$_3$ films grown on BIS-BL exhibit the best crystalline quality, allowing hole-doping to be feasible.

## II. Control and estimation of Ca doping level in Bi$_2$Se$_3$ films

(Ca$_x$Bi$_{1-x}$)$_2$Se$_3$ thin films are grown by codeposition of Ca, Bi, and Se. In order to control Ca doping concentration, x, we need to know the flux of Ca as well as that of Bi during growth: as for Se, its exact flux is not critical as far as it is much higher than that of Bi because extra Se does not stick. The flux of sources is measured with quartz crystal microbalance (QCM). However, unlike Bi and Se, the fluxes for Ca are too low to be reliably measured by QCM. But such low flux values can still be estimated based on QCM measurements done at higher fluxes and the source temperatures as described below.

Based on ideal gas laws, it can be easily shown that

$$\Phi_i = \Phi_j \cdot \frac{P_i}{P_j} \cdot \left(\frac{T_j}{T_i}\right)^{\frac{1}{2}}, \tag{1}$$

where $\Phi_i$ and $\Phi_j$ are source fluxes (/cm$^2$·s), $P_i$ and $P_j$ are vapor pressures (Torr) of the source at the source temperatures, $T_i$ and $T_j$ (K), respectively [3]. The vapor pressure of Ca as a function of temperature is available on line (Table S1), and it turns out that it can be precisely fit by:

$$\log(P) = a_0 + a_1 \cdot T + a_2 \cdot T^2 \tag{2}$$

where $a_0$, $a_1$, and $a_2$ are the fitting parameters. Then, $\Phi_i$ can be expressed as follows.

$$\Phi_i = \Phi_j \cdot \frac{e^{a_0 + a_1 \cdot T_i + a_2 \cdot T_i^2}}{e^{a_0 + a_1 \cdot T_j + a_2 \cdot T_j^2}} \cdot \left(\frac{T_j}{T_i}\right)^{\frac{1}{2}} \tag{3}$$

In order to use Equation (3) for Ca, we fit the vapor pressure and temperature data of Ca in Table S1, and find $a_0$ = - 40.67817, $a_1$ = 0.09757, and $a_2$ = - 6.30447 × 10$^{-5}$. Then, we measure using QCM the Ca flux at a higher source temperature, 580 °C ($T_j$), which turns out to be 5.5 × $10^{13}$ /cm²·s ($\Phi_j$). Substituting the fitting parameters, the source temperature ($T_j$) and the flux ($\Phi_j$) at the higher temperature, Equation (3) becomes a function of $T_i$ only, the source temperature. With the typical flux of Bi in our growth, $2.01 \times 10^{13}$ /cm²·s, we can calculate the Ca concentration in $(Ca_xBi_{1-x})_2Se_3$: $\frac{\Phi_{Ca} \cdot \alpha}{\Phi_{Bi} + \Phi_{Ca} \cdot \alpha} \times 100(\%)$, where $\Phi_{Ca}$ is the Ca flux, $\Phi_{Bi}$ is the Bi flux, and $\alpha$ is a coefficient, which turns out to be 0.36, to adjust the QCM-measured flux to match the doping concentration measured by STM, say, for a 2% Ca-doped $Bi_2Se_3$ film. Using this method, we can estimate the doping concentration at much lower temperatures than can be measured by QCM. The Ca doping levels and the corresponding source temperatures used in our study are shown in Table S2.

# III. Determination of carrier density and mobility in the n-p mixed regime, via two carrier model

Although $R_{xy}(B)$, Hall resistance as a function of magnetic field applied, is nonlinear in n-p mixed regime, we can still extract the sheet carrier density ($n_{2D}$) and the mobility ($\mu$) using a standard multicarrier model, the simplest of which is the two carrier model expressed as:

$$R_{xy}(B) = -\frac{B}{e} \cdot \frac{(n_1\mu_1^2 + n_2\mu_2^2) + (n_1+n_2)\mu_1^2\mu_2^2 B^2}{(n_1\mu_1 + n_2\mu_2)^2 + (n_1+n_2)^2 \mu_1^2\mu_2^2 B^2}, \qquad (4)$$

where $B$ is the magnetic field; $e$ is the electric charge; $n_1$, $n_2$, $\mu_1$, and $\mu_2$ are the sheet carrier densities and mobilities of type 1 and 2, respectively.

Using this two carrier model, we fit the n-p mixed curves and plot the results in Figure S2. Here, we use $n_1$, $n_2$, $\mu_1$, and $\mu_2$ as the fitting parameters with a constraint that the error in longitudinal resistance at $B = 0$ T be less than 3 %. In consequence, we perform "effective three parameter fitting" in $R_{xy}(B)$. As can be seen in Figure S2a, which is for 20 QL Bi$_2$Se$_3$ film with 0.17 % of Ca, the fitting is generally quite good. Figure S2b-i show $n_{2D}$ and $\mu$ vs Ca concentration for four different thicknesses of Bi$_2$Se$_3$ films in the n-p mixed regime, as obtained from the two carrier fitting. Two features stand out from this fitting. First is that in the n-p mixed regime, the carrier density of the p-channel is always much larger than that of the n. Second, the mobility of the p-channel is always much smaller than that of the n. This observation is consistent with the conclusion in the main text, made without relying on the two carrier model: crossing the charge neutral point from n to p-type, the carrier density soars while the mobility plummets.

## IV. Comparison of alkaline earth elements as hole dopants: Mg, Ca, Sr, and Ba

Considering that the nominal valence of Bi in $Bi_2Se_3$ is 3+, any 2+ ions can be a candidate as a hole dopant for $Bi_2Se_3$. Therefore, we compared four alkaline earth elements, Mg, Ca, Sr, and Ba as a hole dopant for their doping efficiency up to 1 % in $Bi_2Se_3$ films: all these films are grown directly on $Al_2O_3$ (0001) substrates and capped by Se. The Hall effect measurements were carried out in an electro-magnet system with base temperature of 5 K and maximum magnetic field of ±0.6 T.

The Hall effect data in Figure S3a-d show that Mg, Ca, and Sr are all similar in their hole doping efficiency, whereas Ba fails to act as a hole dopant. For Mg, Ca, and Sr, the $n_{2D}$ monotonically decreased to similar values at 1% doping: $4.8 \times 10^{12}$ /$cm^2$ for Mg, $4.1 \times 10^{12}$ /$cm^2$ for Ca, and $4.0 \times 10^{12}$ /$cm^2$ for Sr. On the other hand, the $n_{2D}$ for Ba-doped films rose to $1.3 \times 10^{14}$ /$cm^2$ at 1 % doping.

This doping efficiency difference can be partly understood from the comparison of the ionic sizes between the dopants and the Bi ion shown in Figure 3e. The ionic radius of $Ca^{2+}$ (100 pm) is closest to that of $Bi^{3+}$ (103 pm), and that of $Ba^{2+}$ (135 pm) is farthest off [4]. This ionic size comparison suggests that Ca is likely to be the least disruptive when substituted for Bi, and so in combination with the above Hall effect comparison study, we have chosen Ca as the hole dopant. However, as the above Hall effect study shows, Mg and Sr could be also as effective as Ca.

**V. Transport properties of another set of 8 QL Ca-doped Bi$_2$Se$_3$ thin films capped by MoO$_3$ and Se, covering a broader doping range than Figure 4.**

In Figure 4 of the main text, we showed that when the dominant carrier type in 8 QL Ca-doped Bi$_2$Se$_3$ thin films switches from n to p with increasing Ca doping, p-type films still had n-type minority carriers mixed together. In order to see how the transport properties evolve as we move deep into the p-regime, we grew another set of 8 QL Ca-doped Bi$_2$Se$_3$ thin films covering a broader doping range and present their transport properties in Figure S4. These samples are measured at 1.5 K and up to 9 T, instead of at 0.3 K and up to 34.5 T as in Figure 4. The samples that have overlapping doping levels with those in Figure 4 exhibit similar transport properties with slight variations: for example, while the most conducting sample at zero magnetic field in Figure 4 was 0.11%, the most conducting one in Figure S4 is 0.12%. On the p-side, there is n-p mixing at least up to 0.69%, and by the time the samples become pure p-type (0.86% or 1.1%, depending on the view point), the p-type carrier density reaches ~3 × 10$^{12}$ /cm$^2$, which is an order of magnitude larger than the n-type carrier density achieved on the pure n-type samples.

Whenever the Hall resistance is nonlinear, the carrier density cannot be trivially extracted from the Hall resistance curve. One way to get around this problem is to use a multi-carrier model as we did above in section III. However, considering that we cannot be sure about the valid number of conduction channels, such a fitting has its own limitations. In order to avoid any assumptions or biases, we plot two different sets of carrier densities and mobilities evaluated at two different magnetic fields in Figure S4c-f. Generally, we take derivative of $R_{xy}(B)$ to extract $n_{2D}$ from $R_{xy}(B)$ as follows:

$$n_{2D} = -1 \bigg/ \left( e \cdot \left[ \frac{dR_{xy}(B)}{dB} \right] \right). \tag{5}$$

If $R_{xy}(B)$ is linear, $n_{2D}$ is the same regardless of where the derivative is taken in $R_{xy}(B)$. However, if $R_{xy}(B)$ is nonlinear, then $n_{2D}$ value is dependent on where the derivative is taken. Here, we take two different points of the magnetic field for the derivative: 0 T and 9 T. The derivative at 0 T is dominated by $n_{2D}$ of the highest mobility channel. On the other hand, the slope at the high magnetic field limit approaches the net $n_{2D}$ (n-type + p-type): this can be seen from the two carrier model, Equation (4), in section III. Therefore, we take 9 T (the highest field here) as the high magnetic field limit and compare the slopes with the 0 T values. Figure 4Sc-f show that away from the n-p mixed regime, the quantities evaluated at the two magnetic fields are almost the same, whereas they are noticeably different in the n-p mixed regime. The increase in the n-type carrier density before the carrier type changes to p-type is not real and due to the fact that we have not reached high enough magnetic field to fully reveal the low mobility channel (here, p-type). Despite the uncertainty in the carrier densities of the n-p mixed regime, both Figure S4d and f show that the mobility drops sharply as the carrier type changes from n to p-type, which is consistent with what we observed in the main text and in section III.

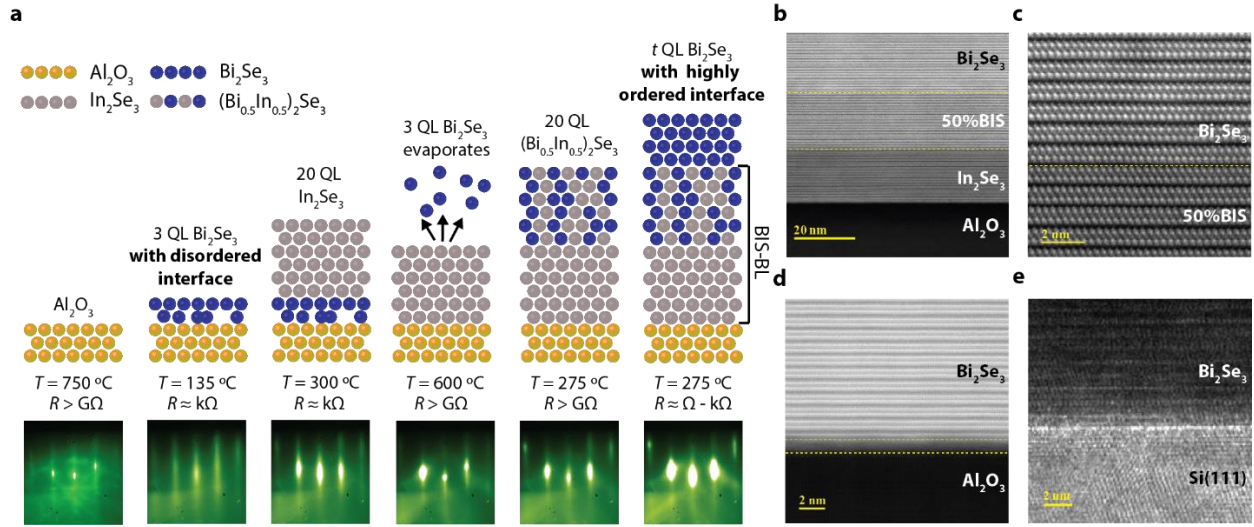

**Figure S1.** Growth process of Bi$_2$Se$_3$ films on the In$_2$Se$_3$-(Bi$_{0.5}$In$_{0.5}$)$_2$Se$_3$ buffer layer (BIS-BL) and comparison with films grown on BIS-BL, Al$_2$O$_3$ (0001) and Si (111), taken from Ref. 2. (a) Cartoon showing each step of the growth process along with the corresponding growth temperature (T), sheet resistance (R) and RHEED images. (b, c) High-angle annular dark-field scanning transmission electron microscopy (HAADF-STEM) images of Bi$_2$Se$_3$ grown on BIS-BL. (c) The image with higher resolution on the interface between Bi$_2$Se$_3$ and BIS shows an atomically sharp interface, while (d) Bi$_2$Se$_3$ grown directly on Al$_2$O$_3$ (0001) has clearly disordered interface, indicated by hazy region in the yellow dotted lines. (e) TEM image of Bi$_2$Se$_3$ grown on Si(111) (from Ref. 5).

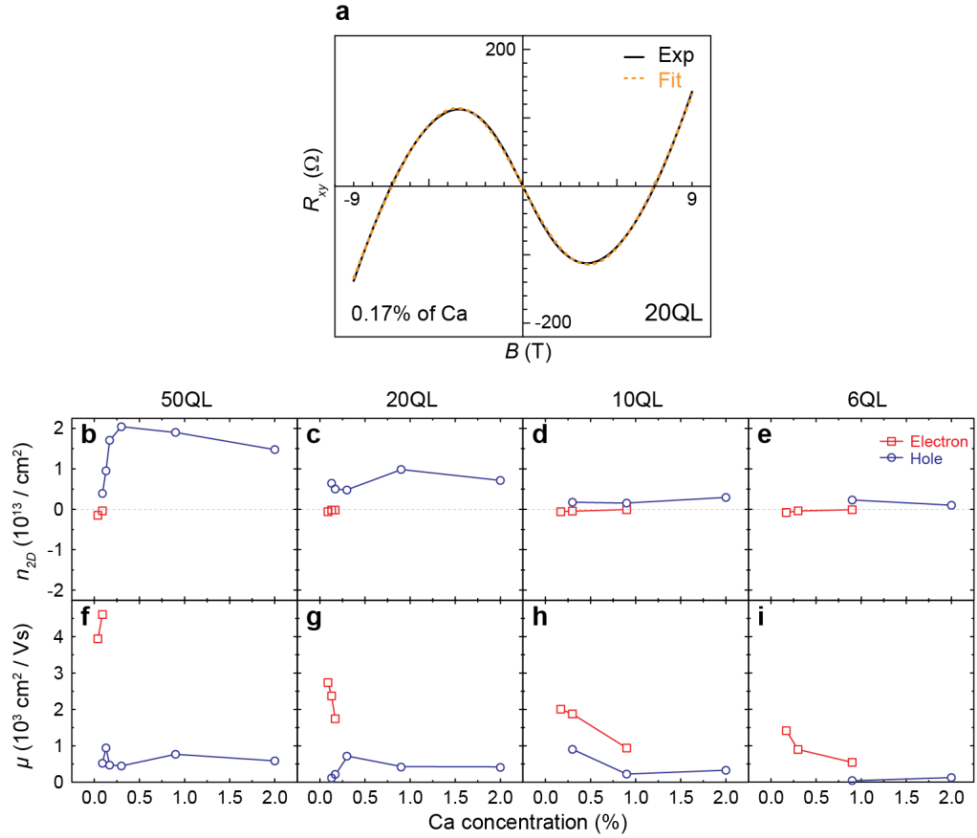

**Figure S2.** Two carrier model fitting of a 20 QL, n-p mixed, Ca-doped Bi$_2$Se$_3$ film grown on BIS-BL and capped by Se and complete data of sheet carrier density and mobility with n-p mix regime. (a) Black solid line shows the actual Hall measurement data, and orange dashed line shows the two carrier model fitting. (b-e) The $n_{2D}$ vs Ca concentration plots with data from the two carrier model fittings of 50 QL (b), 20 QL (c), 10 QL (d), and 6 QL (e), n-p mixed, Ca-doped Bi$_2$Se$_3$ films. (f-i) The $\mu$ vs Ca concentration plots with data from the two carrier model fittings of 50 QL (f), 20 QL (g), 10 QL (h), and 6 QL (i) films.

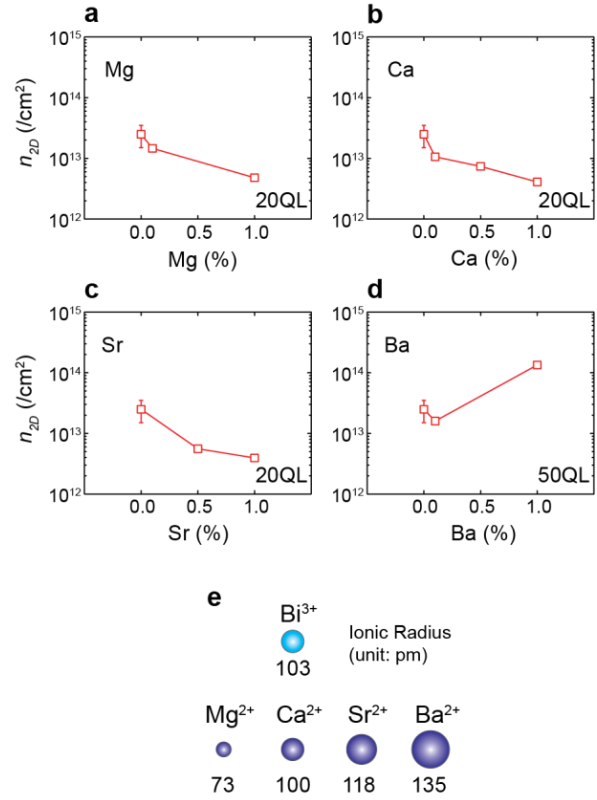

**Figure S3.** Sheet carrier density ($n_{2D}$) vs doping level (measured at 5 K) of alkali earth elements and comparison of ionic radiuses. (a-d) Mg (a), Ca (b) and Sr (c) doping in 20 QL, and Ba (d) doping in 50 QL of $Bi_2Se_3$ films grown directly on $Al_2O_3$ (0001) substrates and capped by Se. The $n_{2D}$ of pure $Bi_2Se_3$ films, $2.5 \times 10^{13}$ /cm$^2$, is included with an error bar of $\pm\ 1.0 \times 10^{13}$ /cm$^2$ in (a-d). (e) Comparison of ionic radiuses of $Bi^{3+}$ and dopants, $Mg^{2+}$, $Ca^{2+}$, $Sr^{2+}$, and $Ba^{2+}$.

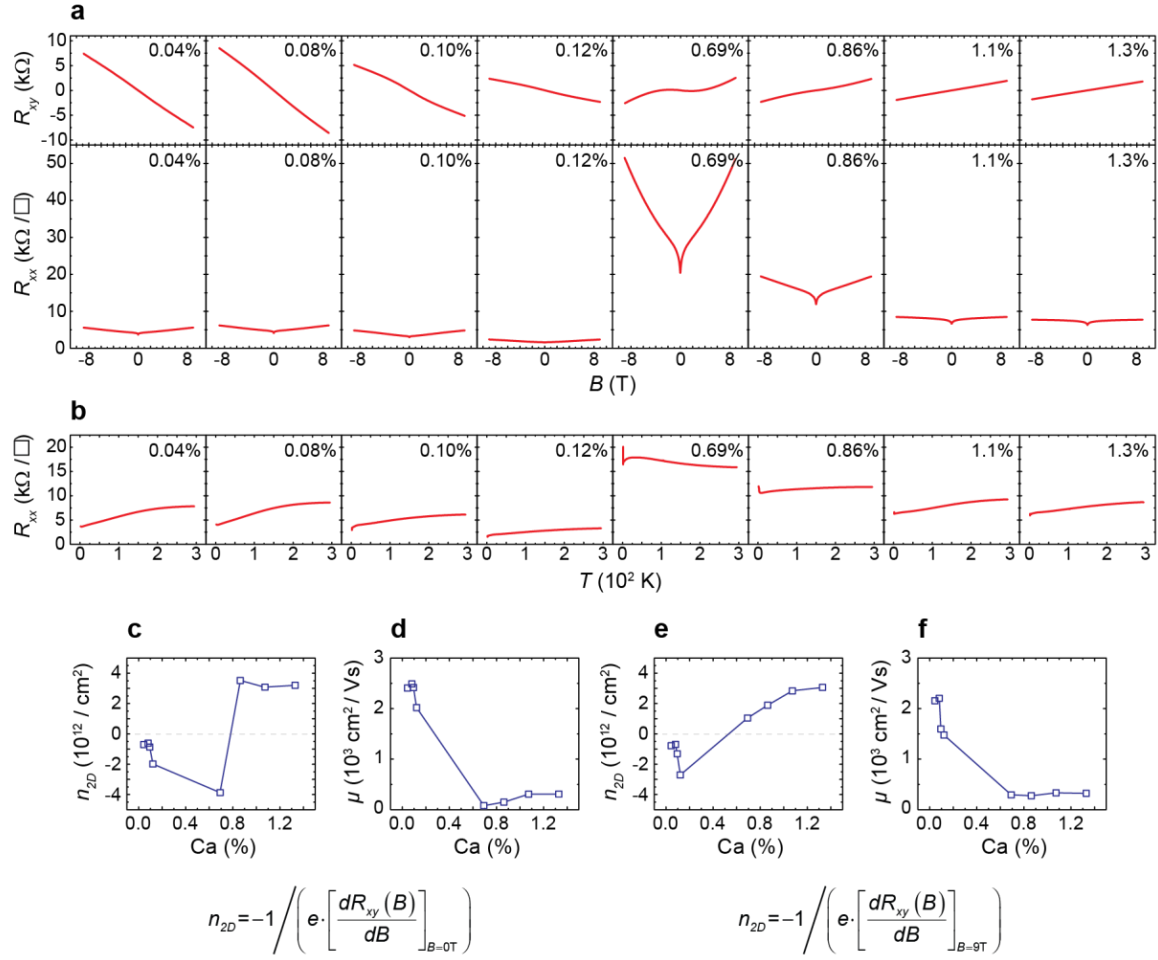

**Figure S4.** Transport properties of another set of 8 QL Ca-doped $Bi_2Se_3$ thin films, covering a broader doping range than Figure 4 of the main text, measured at or down to 1.5 K. (a) Ca doping dependence of $R_{xy}$ and $R_{xx}$ vs $B$. (b) Ca doping dependence of $R_{xx}$ vs $T$. (c, d) $n_{2D}$ (c) and $\mu$ (d) vs Ca concentration using $(dR_{xy}/dB)_{B=0T}$ to extract $n_{2D}$. (e, f) $n_{2D}$ (e) and $\mu$ (f) vs Ca concentration using $(dR_{xy}/dB)_{B=9T}$ to extract $n_{2D}$.

**Table S1.** Vapor pressures at selected source temperatures of Ca. The data are available from Veeco.

| Vapor pressure (Torr) | $10^{-7}$ | $10^{-6}$ | $10^{-5}$ | $10^{-4}$ | $10^{-3}$ |
|---|---|---|---|---|---|
| Temperature (°C) | 317 | 357 | 405 | 459 | 522 |

**Table S2.** Ca doping level in $Bi_2Se_3$ films controlled by the source temperature. The flux is estimated by the expression, Equation (3), and the doping level is calculated by $\Phi_{Ca} \cdot \alpha / (\Phi_{Bi} + \Phi_{Ca} \cdot \alpha) \times 100$ (%) with $\alpha = 0.36$, $\Phi_{Bi} = 2.01 \times 10^{13}$ /cm²·s.

| Temperature (°C) | Flux ($\Phi_{Ca}$, $10^{10}$ /cm²·s) | Doping level (%) |
| --- | --- | --- |
| 355 | 2.5 | 0.04 |
| 365 | 4.1 | 0.07 |
| 367 | 4.5 | 0.08 |
| 368 | 4.8 | 0.09 |
| 370 | 5.3 | 0.10 |
| 373 | 6.1 | 0.11 |
| 375 | 6.8 | 0.12 |
| 376 | 7.1 | 0.13 |
| 378 | 7.9 | 0.14 |
| 382 | 9.6 | 0.17 |
| 394 | 17.0 | 0.30 |
| 412 | 38.9 | 0.69 |
| 417 | 48.6 | 0.86 |
| 422 | 60.5 | 1.07 |
| 427 | 75.1 | 1.33 |
| 437 | 114.8 | 2.02 |
| 457 | 258.2 | 4.42 |